\documentclass[aps,prl,superscriptaddress,amsmath,groupedaddress,twocolumn]{revtex4}

\usepackage{color,hangcaption,hhline,psfrag,rotating,amssymb}
\usepackage[hang,nooneline]{subfigure}
\usepackage{dcolumn}

\begin{document}
\title{A Prediction of the $B^*_c$ mass in full lattice QCD}

\author{E. B. Gregory}
\email[]{e.gregory@physics.gla.ac.uk}
\affiliation{Department of Physics and Astronomy, University of Glasgow, Glasgow, G12 8QQ, UK}
\author{C. T. H. Davies}
\affiliation{Department of Physics and Astronomy, University of Glasgow, Glasgow, G12 8QQ, UK}
\author{E. Follana}
\affiliation{Departamento de F\'{\i}sica Te\'{o}rica, Universidad de Zaragoza, E-50009 Zaragoza, Spain}
\author{E. Gamiz}
\affiliation{Dept. of Physics, University of Illinois, 1110 West Green Street, Urbana, Illinois 61801, USA}
\author{I.~D.~Kendall}
\affiliation{Department of Physics and Astronomy, University of Glasgow, Glasgow, G12 8QQ, UK}
\author{G. P. Lepage}
\affiliation{Laboratory of Elementary-Particle Physics, Cornell University, Ithaca, New York 14853, USA}
\author{H. Na}
\affiliation{Department of Physics, The Ohio State University, Columbus, Ohio 43210, USA}
\author{J. Shigemitsu}
\affiliation{Department of Physics, The Ohio State University, Columbus, Ohio 43210, USA}
\author{K. Y. Wong}
\affiliation{Department of Physics and Astronomy, University of Glasgow, Glasgow, G12 8QQ, UK}


\collaboration{HPQCD collaboration}
\homepage{http://www.physics.gla.ac.uk/HPQCD}
\noaffiliation

\date{\today}

\begin{abstract}
By using the Highly Improved Staggered Quark formalism to handle charm, strange
and light valence quarks in full lattice QCD, and NRQCD to handle bottom 
valence quarks we are able to determine accurately ratios of the $B$ meson 
vector-pseudoscalar mass splittings, in particular, $(m(B^*_c) - m(B_c))/(m(B^*_s) - m(B_s))$. We find 
this ratio to be 1.15(15), showing the `light' quark mass dependence of 
this splitting to be very small. Hence we predict $m(B_c^*) = 6.330(7)(2)(6)$ GeV where 
the first two errors are from the lattice calculation and the third from existing experiment. 
This is the most accurate prediction of a gold-plated hadron mass from lattice QCD to date. 
\end{abstract}


\maketitle

\section{Introduction}
Particle physicists now have long familiarity with 
the low-lying spectrum of $b\overline{b}$ (Upsilon) and 
$c\overline{c}$ (psi) mesons but they nevertheless continue
to provide a very important testing ground for our understanding
of strong interaction physics. The similar $b\overline{c}$ ($B_c$) system, 
on the other hand, 
is largely unexplored territory so predictions of the meson masses 
are very valuable. These predictions need to be as accurate as 
possible (and with an error budget) to provide stringent tests 
of QCD.
Lattice QCD is clearly one of the best ways to do this, now that 
accurate calculations including the full effect of 
$u$, $d$, and $s$ sea quarks inside hadrons are possible~\cite{ratiopaper}. 
It has already provided successful predictions of the pseudoscalar $\eta_b$ mass~\cite{alan} 
(with a 14 MeV error)
and the $B_c$ mass~\cite{allison} (with a 20 MeV error), both subsequently seen 
by experiment. Here we give a prediction 
for the vector $B_c^*$ mass through the mass difference between the $B_c^*$ and 
the $B_c$. 

Mesons composed of valence heavy ($b$ and $c$) quarks are 
relatively simple because they are nonrelativistic systems 
and a potential model may be 
expected to work reasonably well (see, for example,~\cite{kwong-rosner, equigg, gershtein}). 
This is especially true for the 
$\Upsilon$ system where $v_b^2 \approx 0.1$ (in units of $c^2$). It is less true for 
charmonium where $v_c^2 \approx 0.3$ and so relativistic corrections 
are much larger there. The ground state hyperfine 
(vector-pseudoscalar) mass splitting is such a correction, but 
is given in leading order perturbation theory by a simple
formula since the $\vec{S}.\vec{S}$ potential is proportional 
to $\delta^3(\vec{r})$.
\begin{equation}
\Delta M = \frac{32 \pi \alpha_s |\psi(0)|^2 }{9 m_1 m_2},
\label{eq:hyp}
\end{equation}
where $m_1$ and $m_2$ are the masses of the quark and antiquark and 
$\psi(0)$ is the wavefunction at the origin from the potential model. 
Using this formula to calculate the splitting will have a 
systematic error at $\cal{O}$$(v^2)$ i.e. 30\% in $c\overline{c}$, 
10\% in $b\overline{b}$ and 20\% in $b\overline{c}$. However, 
the $c\overline{c}$ hyperfine splitting has been used in the past 
to fix the effective value of $\alpha_s$ in eq.~\ref{eq:hyp} 
and then that 30\% error affects all subsequent calculations. 
A larger problem, perhaps, is the variation in 
results between different potential models tuned to the spin-independent 
spectrum. This is because that spectrum does not in practice constrain the 
wavefunction at the origin at all strongly. 
The mass splitting between $B_c^*$ and $B_c$ can vary in the 
range 40-90 MeV~\cite{kwong-rosner, equigg, gershtein} between different potentials, which makes it hard to decide a `central 
value' and error budget.   

The reduced mass in the $B_c$ system is roughly one half that of 
$b\overline{b}$ and 1.5 times that of the $c\overline{c}$. 
Then $v_b^2 \approx 0.05$ in $B_c$ but 
$v_c^2 \approx 0.4-0.5$, which makes a nonrelativistic treatment
worse in principle than for charmonium. An alternative approach is 
to treat the $B_c$ as a `heavy-light' system using ideas 
from HQET but, for example, it is difficult to estimate the light quark mass 
dependence of the $1/m_Q$ operator giving rise to the hyperfine splitting, 
limiting again the accuracy in the prediction.   

Lattice QCD, on the other hand, can provide very stringent tests 
of QCD from the hadron spectrum, 
in which all sources of 
systematic error can now be tested and 
quantified~\cite{ratiopaper}. The only parameters are those of QCD itself 
(a quark mass for every flavor and a coupling constant) and 
impressively accurate results in agreement with experiment 
can be produced for the whole range of gold-plated hadron 
masses known experimentally.  
Our previous prediction of the $B_c$ mass~\cite{allison} dates 
from the relatively early days of full lattice QCD 
calculations and is now being improved. We 
have since developed a much more accurate method for handling charm 
quarks within lattice QCD and that has enabled a
determination of the $B_c$ mass with smaller systematic errors~\cite{usinprep}. 
This method also allows an accurate prediction of the
$B_c^*$ and we describe that calculation here. 

\section{LATTICE QCD CALCULATION}

From above it is clear that an optimal 
lattice QCD approach to the $B_c$ is to combine a nonrelativistic method for 
the $b$ quark with a relativistic one for $c$. 
Here we use Lattice NRQCD for the $b$, 
developed over many
years~\cite{thackerlepage, nakhleh, oldups} to 
provide accurate bottomonium spectroscopy~\cite{alan} by including 
spin-independent terms through $\cal{O}$$(v_b^4)$ and leading spin-dependent terms
with discretisation corrections through $\cal{O}$$(a^2)$. 
For the $c$ quark we use Highly Improved Staggered Quarks (HISQ)~\cite{hisq}, 
a fully relativistic discretisation of the Dirac action which 
is accurate enough to handle $c$ quarks because it is fully 
improved through $\cal{O}$$(a^2)$ and also has the leading 
$(m_ca)^4$ errors removed. This enables us to use the 
same lattice QCD action for charm, strange and light quarks (we take $m_u = m_d$). 
This approach, as we shall see, enables us 
to cancel some systematic errors between the 
$B_c$ system and the $B_s$ system  and obtain the hyperfine 
splitting in the $B_c$ as a multiple of the experimentally 
known splitting~\cite{pdg09} in the $B_s$ system. 

\begin{table}
\begin{tabular}{lllllllll}
\hline
\hline
Set & $\beta$ & $r_1/a$ & $au_0m_{0l}^{asq}$ & $au_0m_{0s}^{asq}$ & $L/a$ & $T/a$ & $N_{conf}\times N_{t}$ \\
\hline
1 & 6.572 & 2.152(5) & 0.0097 & 0.0484 & 16 & 48 & $624 \times 2$\\
2 & 6.586 & 2.138(4) & 0.0194 & 0.0484 & 16 & 48 & $628 \times 2$\\
\hline
3 & 6.760 & 2.647(3) & 0.005 & 0.05 & 24 & 64 & $507 \times 2$ \\
4 & 6.760 & 2.618(3) & 0.01 & 0.05 & 20 & 64 & $589 \times 2 $ \\
\hline
5 & 7.090 & 3.699(3) & 0.0062 & 0.031 & 28 & 96 & $530 \times 4$ \\
\hline
\hline
\end{tabular}
\caption{\label{tab:params}Ensembles (sets) of MILC configurations used with gauge coupling $\beta$, 
size $L^3 \times T$ and sea 
masses ($\times$ tadpole parameter, $u_0$) 
$m_{0l}^{asq}$ and $m_{0s}^{asq}$. 
The lattice spacing values in units of $r_1$ after `smoothing'
are given in column 3~\cite{milcreview}. 
Column 8
gives the number of configurations and time sources per configuration 
that we used for calculating correlators. On set 5 only half the 
number were used for light quarks.}
\end{table}

\begin{table}
\begin{tabular}{lllllll}
\hline
\hline
Set & $aM_{b}^0$ & $u_{0L}$ & $am_{0c}^{hisq}$ & $1+\epsilon$ & $am_{0s}^{hisq}$ & $am_{0l}^{hisq}$  \\
\hline
1 & 3.4 & 0.8218 & 0.85 & 0.66 & 0.066 & 0.0132 \\
2 & 3.4 & 0.8225 & 0.85 & 0.66 & 0.066 & 0.0264 \\
\hline
3 & 2.8 & 0.8362 & 0.65 & 0.79 & 0.0537 & 0.0067 \\
4 & 2.8 & 0.8359 & 0.66 & 0.79 & 0.05465 & 0.01365 \\
\hline
5 & 1.95 & 0.8541 & 0.43 & 0.885 & 0.0366 & 0.00705 \\
\hline
\hline
\end{tabular}
\caption{\label{tab:valparams} Parameters for the valence quarks. 
$aM_{b}^0$ is the $b$ quark mass in NRQCD, and $u_{0L}$ is the tadpole-improvement 
parameter used there~\cite{alan}. We use stability parameter~\cite{alan} $n$ = 4
everywhere. 
Since NRQCD quarks propagate in one direction in time only we improve 
statistics by generating propagators both forwards in time (for $T/2$ 
time units) and backwards 
in time from each source. 
Columns 4, 6 and 7 give the charm, strange and light bare quark masses 
for the HISQ action. $1+\epsilon$ is the coefficient of the Naik term
in the charm case~\cite{hisq}. 
}
\end{table}

\begin{figure}[ht]
\begin{center}
\includegraphics[width=7.0cm]{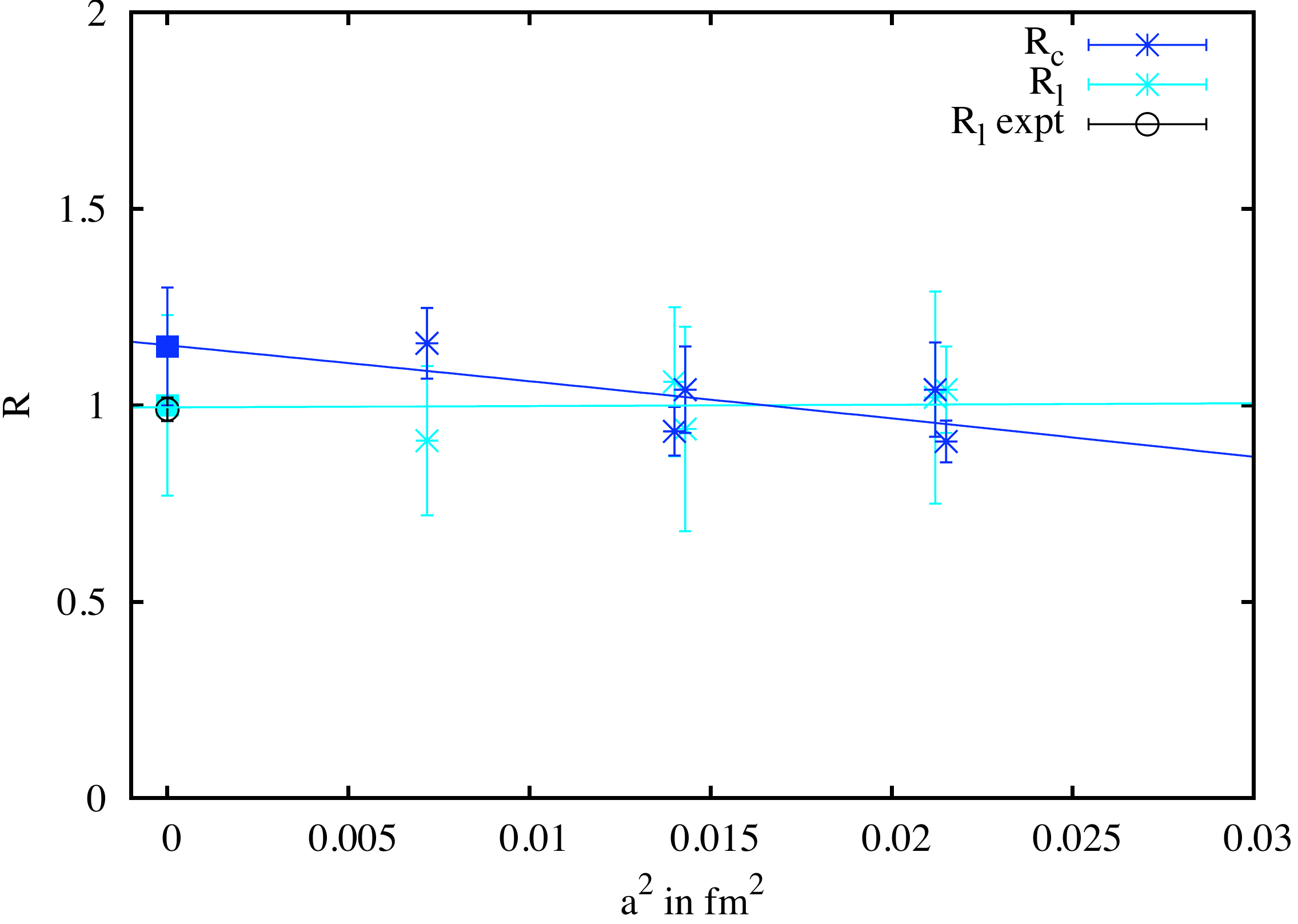}
\end{center}
\caption{\label{fig:bc_split}The ratio $R_c$ of the $B_c$ and $B_s$ hyperfine splittings 
(eq.~\ref{split_rat})
as a function of lattice spacing, $a$, from full lattice QCD. Our continuum extrapolation 
is also given and the result at $a=0$. The lighter points and line give the equivalent 
points for $R_l$ along with the experimental value~\cite{pdg09}.} 
\end{figure}

We work with five ensembles of gluon field configurations provided 
by the MILC collaboration. These include the full effect of 
$u$, $d$ and $s$ sea quarks using the improved staggered (asqtad)
formalism and are available with large spatial volumes ($>(2.4{\mathrm{fm}})^3$)
and at multiple values of the light sea masses (using $m_u=m_d$) for 
a large range of lattice spacing values. 
We use configurations at three values of $a$
between 0.15 fm and 0.09 fm with parameters as listed 
in Table~\ref{tab:params}.  
On each configuration in the ensemble we generate $b$ quark propagators
using NRQCD and $c$, $s$ and $l$ quark propagators using HISQ. 
The parameters of the valence quarks are given in Table~\ref{tab:valparams}. 
The $b$ quark mass is tuned to give the correct $\Upsilon$ mass~\cite{r1paper}
and the charm, strange and light masses are taken from~\cite{fds}. 

The $b$ quark is then combined in turn with each of the other 
three with appropriate spin matrices to make pseudoscalar or vector mesons. 
To increase statistics we generate propagators from sources at several 
different timeslices per configuration (see Table~\ref{tab:params}). 
We also use a random wall source for the quarks~\cite{fds}, taken 
as a set of U(1) random numbers at each point on the source 
time slice. 
This mimics multiple sources across a timeslice when 
the propagators are paired up, improving statistics further. 
For the NRQCD propagators, as well as a local source, we also need `smeared' sources~\cite{alan} 
chosen to improve the overlap with the ground state in the meson 
correlator. 
Exponentially growing noise is a problem in the $B$ system 
(particularly as the lighter quark mass becomes small)
and smearing enables us to extract an accurate ground state energy 
from the correlator at smaller time separation from the source than 
otherwise~\cite{eglat08}. 
We use a Gaussian form 
for the smearing function with radius 2$a$ and 4$a$. These various sources for the NRQCD 
quark must be combined with the random wall described above.  
In addition the NRQCD quark source must now include the matrix 
that converts spinless staggered quarks into naive quarks
for combination with 2-spin NRQCD quarks in an adaption~\cite{eglat08, usinprep} of the 
standard method of combining heavy quarks with staggered quarks~\cite{wingate}. 

\begin{figure*}[htb]
\begin{center}
\includegraphics[width=13.0cm]{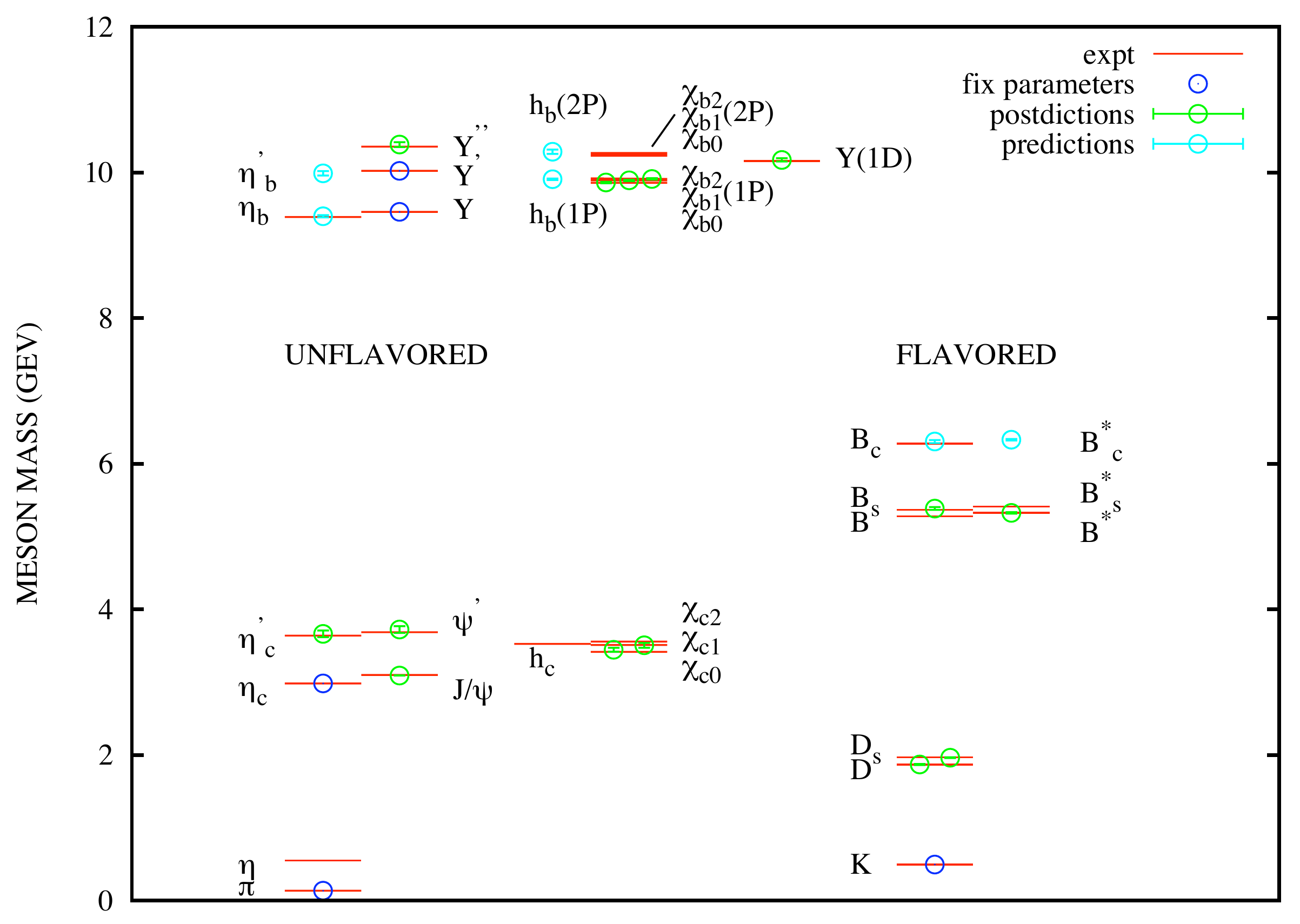}
\end{center}
\caption{The spectrum of `gold-plated' mesons from HPQCD calculations. Results are divided into those 
used to fix the parameters of QCD (4 quark masses and a coupling constant); 
those which are postdictions~\cite{alan, hisq, fds} and those, 
like the $B_c^*$ described here, which are predictions~\cite{alan, allison}.  }
\label{fig:goldmesonplot}
\end{figure*}

We fit our $3\times3$ matrix~\cite{fnmatrix} of $B$ meson correlators 
using the standard Bayesian method~\cite{gplbayes} 
to a sum of exponentials, including oscillating parity partner 
states as:
\begin{eqnarray}
C_{\rm B}(i,j;t-t_0)&=& \sum^{N_{\exp}-1}_{k=0} a_{i,k}a^*_{j,k}e^{-E_k(t-t_0)
} \\
&+& \sum^{N_{\exp}-1}_{k^\prime=0}   b_{i,k^\prime}b^*_{j,k^\prime}(-1)^{(t-t_0)
}e
^{-E^\prime_{k^\prime}(t-t_0)}, \nonumber
\label{eq:fit}
\end{eqnarray}
where $i,j$ index different smearing radii. 
We look for stability in the fits and their errors as a function of $N_{\exp}$ 
for ground state energies, $E_0$. 

$E_0$ is not the meson mass but contains an 
energy shift due to the non-relativistic treatment 
of the $b$ quarks~\cite{thackerlepage, nakhleh, oldups}. 
The shift cancels in the mass difference between states with 
the same NRQCD quark content. 
Thus the $B_q^*-B_q$ splitting is obtained directly
from $\Delta_q=E_0(B_q^*)-E_0(B_q)$. 
Because errors are strongly correlated between similar 
quantities calculated on the same ensembles
we fit $B_q$ and $B_q^*$ correlators 
simultaneously to the form above and determine $\Delta_q$ directly 
from the fit.  Values are given
in Table~\ref{tab:results} for $q=l,s,c$.  

\begin{table}[h]
\begin{tabular}{lccccc}
\hline
\hline
Set    &  $\Delta_l$ & $\Delta_s$ & 
$\Delta_c$ &$R_l$ & $R_c$ \\
\hline
1      & 0.0318(78) & 0.0311(37) & 0.0324(2)   & 1.02(27)   & 1.04(12) \\
2      & 0.0374(35) & 0.0359(21)   & 0.0326(3)   & 1.04(11) & 0.908(53)\\
\hline
3      & 0.0306(54) & 0.0287(19)   & 0.0268(2)   & 1.06(19) & 0.934(62) \\
4      & 0.0245(68) & 0.0261(27)   & 0.0271(4)   & 0.94(26) & 1.04(11) \\

\hline
5      & 0.0177(35)  & 0.0190(14)   & 0.0220(6)  & 0.93(19) & 1.158(91)\\
\hline
$a=0$ & & & & 1.00(23) & 1.15(15)\\
\hline
\end{tabular}
\caption{\label{tab:results} Results for the mass differences between 
vector and pseudoscalar B mesons for different light quark content on 
different MILC ensembles. 
$\Delta_q = E_0(B_q^*)-E_0(B_q)$. $R_q$ is the ratio $\Delta_q/\Delta_s$ and 
the result of extrapolating $R_c$ and $R_l$ to $a=0$ is also given. }
\end{table}

The terms from the HISQ action that contribute to the hyperfine
splitting are hidden inside the discretisation of the Dirac covariant derivative. 
Because HISQ is a relativistic action, these terms 
will automatically be correct in the $a \rightarrow 0$ limit. 
The spin-dependent term in the NRQCD action that gives rise 
to the hyperfine splitting can instead be explicitly pinpointed as
the $\vec{\sigma}\cdot\vec{B}$ term~\cite{nakhleh}. This term has the correct 
tree-level coefficient to match full QCD at $\cal{O}$$(v_b^4)$ but 
radiative corrections beyond this have not been included. 
Hence the normalisation of this term, and the normalisation of 
the hyperfine splitting, have an uncertainty of $\cal{O}$$(\alpha_s)$ 
($\approx$ 20\%). This uncertainty is part of the NRQCD action 
and hence the same uncertainty appears regardless of which light 
quark is combined with the $b$ quark and cancels 
in ratios of hyperfine splittings.  
In Table~\ref{tab:results} we also give values for 
\begin{equation}
\label{split_rat}
R_c=\frac{\Delta_c}{\Delta_s}=\frac{E_0(B^*_c)- E_0(B_c)}{E_0(B^*_s)- E_0(B_s)}.
\end{equation}
and the corresponding quantity, $R_l$, for $u/d$ quarks. 
On sets 1-4 $R_l$ is given directly by a joint fit to 
$B_s$ and $B_l$ correlators. 
All $R_l$ values agree with 1 within 30\% errors. 

Figure \ref{fig:bc_split} shows $R_c$ as a function of 
lattice spacing. There is little dependence on the 
light quark mass, since neither the 
$B_c$ nor the $B_s$ contain valence light quarks and 
we do not expect strong sensitivity to the sea content. 
Lattice spacing dependence is mild --- the dashed line 
shows an extrapolation to the continuum 
limit at $a=0$ that can be compared to experiment. 
The extrapolation includes $a^2$ and $a^4$ terms and 
allows for linear dependence on sea quark masses. 
In that limit we find $R=1.15(15)$. This, along with the 
results for $R_l$ show, somewhat 
surprisingly, that the hyperfine splitting varies hardly 
at all with the mass of the lighter quark in the $B$ system, up 
to and including charm. 

The result is backed up by the existing experimental 
results on heavy-light and heavy-strange mesons. In the $D$ system 
the hyperfine splittings differ by only 2\% between  
the $D_s$ and the $D_d$. Some of this difference may in fact be 
a result of coupled-channel effects since the $D_d^*$ is just 
above threshold for the decay to $D\pi$, whereas the $D_s^*$ has 
only the OZI-disfavoured decay mode $D_s\pi$ available. 
The experimental situation is less clear in the $B$ sector since 
some experimental results favour a $B_s^*-B_s$ splitting 
very close to the $B^*-B$ (not yet differentiated into charged 
and neutral modes) and others favour a somewhat larger splitting~\cite{pdg09}. 
We use the PDG average value of 46.1(1.5) MeV~\cite{pdg09} for the $B_s^*-B_s$ splitting
because, in keeping with our 
result and indications from the $D$ sector, 
this is closer to the $B^*-B$ splitting than the PDG fit value of 49.0(1.5) MeV~\cite{pdg09}. 

Our result for $R$ gives
53(7) MeV for the $B_c^*-B_c$ splitting, where the error is 
statistical only. 
Additional systematic errors come from relativistic corrections 
to the $\vec{\sigma}\cdot\vec{B}$ term in the NRQCD action~\cite{nakhleh}. 
We can estimate the size of these from the size of $v_b^2$ 
in the $B_c$ (0.05) and the $B_s$
(=($\Lambda_{QCD}/m_b)^2$ = 0.01). The cancellation between them leads to a
4\% systematic error. Any mistuning of the $b$ quark mass 
will cancel in $R$, and small mistunings of the 
$s$ and $c$ quark masses lead to a negligible error.  
Electromagnetic hyperfine effects missing from our calculation should 
also be negligible (less than 1\%). 

\section{Conclusions}

Adding our value for the $B_c^*-B_c$ splitting to the experimental 
mass for the $B_c$ gives the mass of the $B_c^*$ as 
6.330(7)(2)(6) GeV where the first two errors are from the lattice 
QCD calculation - statistics and systematics respectively - and 
the third error is from experiment for the $B_c$ and the $B_s^*$. 
The relatively small value of the $B_c^*-B_c$ splitting will 
make it challenging to find the $B_c^*$ from its decay to 
$B_c\gamma$. 

The absence of strong dependence of the hyperfine splitting 
on the mass of the lighter quark in the $B$ system is 
an interesting result, which has implications for other spin-dependent 
splittings in the $B_c$ system. In HQET language it says that matrix 
elements of the hyperfine operator are insensitive to the light quark 
mass, up to and including charm. In constituent quark model 
language, using a formula for the hyperfine splitting akin to that in eq.~\ref{eq:hyp}, 
the result implies that $|\psi(0)|^2$
varies as $m_q$ to cancel the $m_q$ in the denominator~\cite{kwong-rosner}. 
The amplitude $a_{loc,0}$ from the fit in eq.~\ref{eq:fit}
is proportional to $\psi(0)$ in a nonrelativistic approach. 
This does show significant dependence on the light quark 
mass in the $B$. For example, $a_{loc,0}(B_c)/a_{loc,0}(B_s) \approx 2$~\cite{usinprep}. 

Finally, in Figure~\ref{fig:goldmesonplot} we summarise the 
current status of the gold-plated meson spectrum as determined 
from lattice QCD, highlighting those meson masses which 
have been made as predictions ahead of experiment. 
The result here is the most accurate prediction to date. 

{\bf{Acknowledgements}} We are grateful to MILC for their 
configurations, to C. McNeile and J. Koponen for useful discussions 
and to STFC, MICINN, NSF and DoE for support. 
Computing was done on USQCD's Fermilab cluster and at the Ohio Supercomputer 
Centre.


\begin{thebibliography}{99}
\bibitem{fnmatrix} We use $1 \times 1$ (local-local) fits for $B_c$ on sets 1 and 2.
\end{thebibliography}
\end{document}